\documentclass[12pt]{article}
\usepackage{graphicx}
\begin{document}
\bigskip

\centerline{\bf Transfer in multi-theme opinion dynamics of Deffuant et al}

\bigskip

Dietrich Stauffer* and Janusz A. Ho{\l}yst 

\bigskip
\noindent
Faculty of Physics, Warsaw University of Technology, \\
Koszykowa 75, PL-00--662 Warsaw, Poland
\bigskip

\noindent
*Visiting from: Institute for Theoretical Physics, Cologne University, \\
D-50923 K\"oln, Euroland,

\bigskip

\noindent
Abstract: {\small Monte Carlo simulations mix the opinion dynamics of Deffuant
et al with the cultural transfer model of Axelrod, using ten discrete 
possible opinions on ten different themes. As Jacobmeier's simulations of
the pure Deffuant case, people preferably agree on nearly all or nearly
no theme. 
}


\bigskip
\bigskip
Opinion dynamics has been simulated with various models, including the 
``negotiators'' of Deffuant et al \cite{deffuant}. Such opinions can be 
expressed on a variety of themes, as in the Axelrod model of cultural transfer
\cite{axelrod}; see \cite{mallorca} for a review. Deffuant et al let people
negotiate with each other repeatedly until they may have reached an agreement.
Axelrod lets everyone accept fully the opinion of somebody else on one theme.
The set of people with whom we talk is restricted to those who do not differ
too much from our own opinion. 

The present work combines the negotiation iterations of Deffuant et al with the 
transfer process of Axelrod, that means at each meeting of two people
they follow Axelrod with probability $p$ and Deffuant et al with probability
$1-p$. Since people usually are not located on regular lattices, we let them
sit on directed Barab\'asi-Albert networks \cite{barabasi}, which may 
\cite{barabasi} or may not \cite{schnegg} be a good approximation for social 
networks \cite{wassermann}. Then for Deffuant negotiators, Jacobmeier 
\cite{jacob,newbook} found for 
ten themes and ten possible opinions for each theme, that usually two people
agree on very few or on nearly all themes, but rarely on half of the themes.
We now check how this result for $p=0$ is affected if we introduce Axelrod
transfer, i.e. for $p > 0$.

\begin{figure}[hbt]
\begin{center}
\includegraphics[angle=-90,scale=0.5]{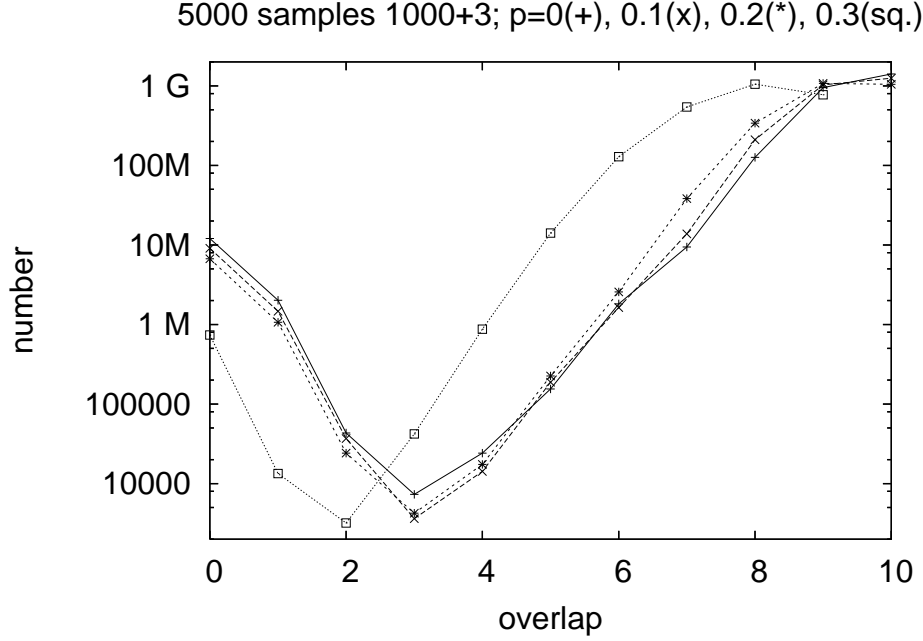}
\end{center}
\caption{Histogram for the overlap in the opinions of all pairs of people;
this overlap can vary from zero (agreement on none of the ten themes) to ten
(agreement in all themes.) The plus signs correspond to the previous case
of $p=0$ \cite{jacob,newbook}, the other signs to nonzero $p$ (admixture
of Axelrod transfer); $\Delta = 30$.
}
\end{figure}  
%'axeldeff17.fig' w lp, 'axeldeff18.fig' w lp, 'axeldeff20.fig' w lp, 'axeldeff22.fig' w lp

\begin{figure}[hbt]
\begin{center}
\includegraphics[angle=-90,scale=0.5]{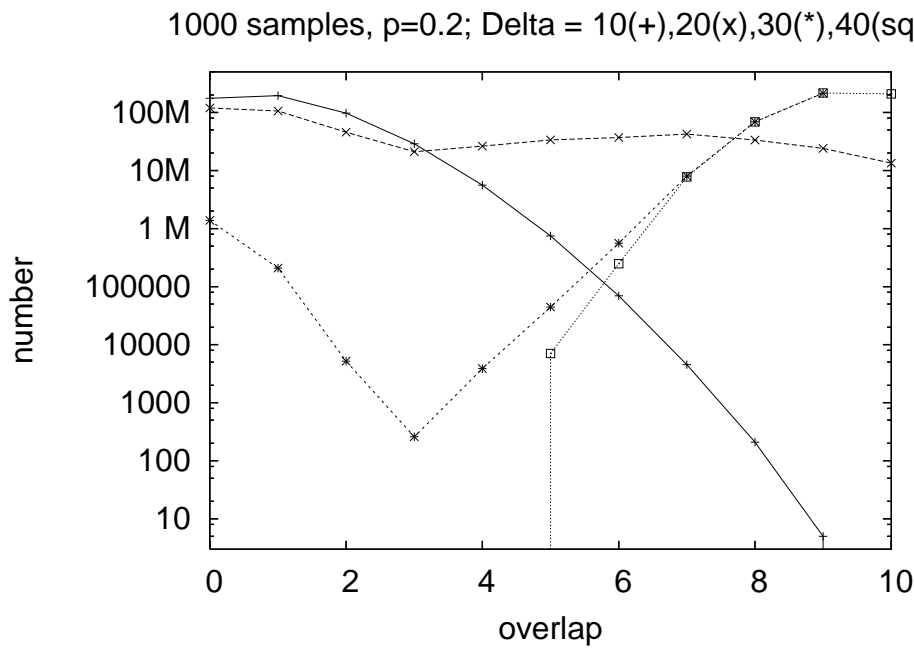}
\end{center}
\caption{Variation of overlap with $\Delta$; 1000+3 sites.
}
\end{figure}  
% axeldeff23*.fig, *=1,2,3,4

\begin{figure}[hbt] 
\begin{center}
\includegraphics[angle=-90,scale=0.32]{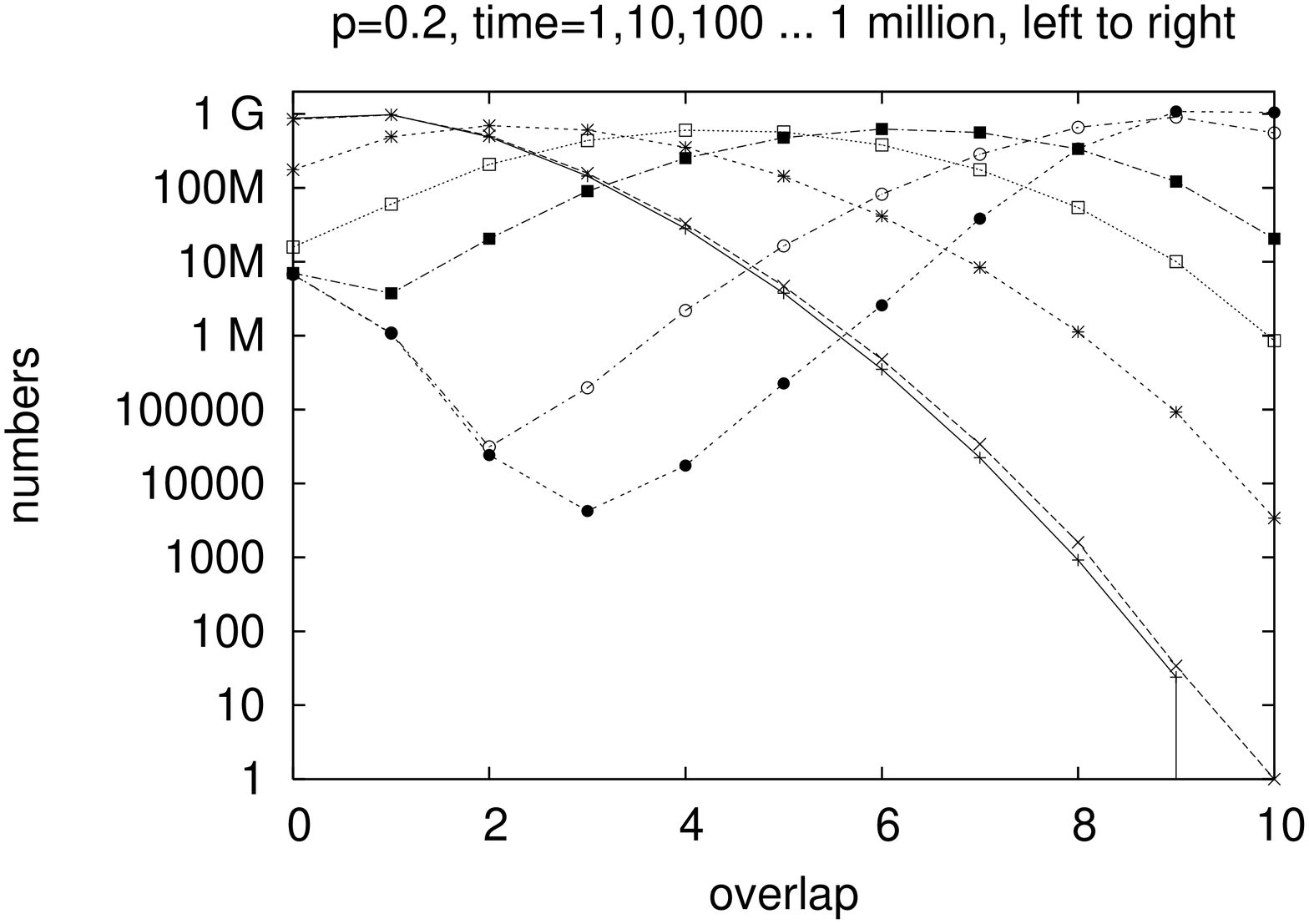}
\includegraphics[angle=-90,scale=0.32]{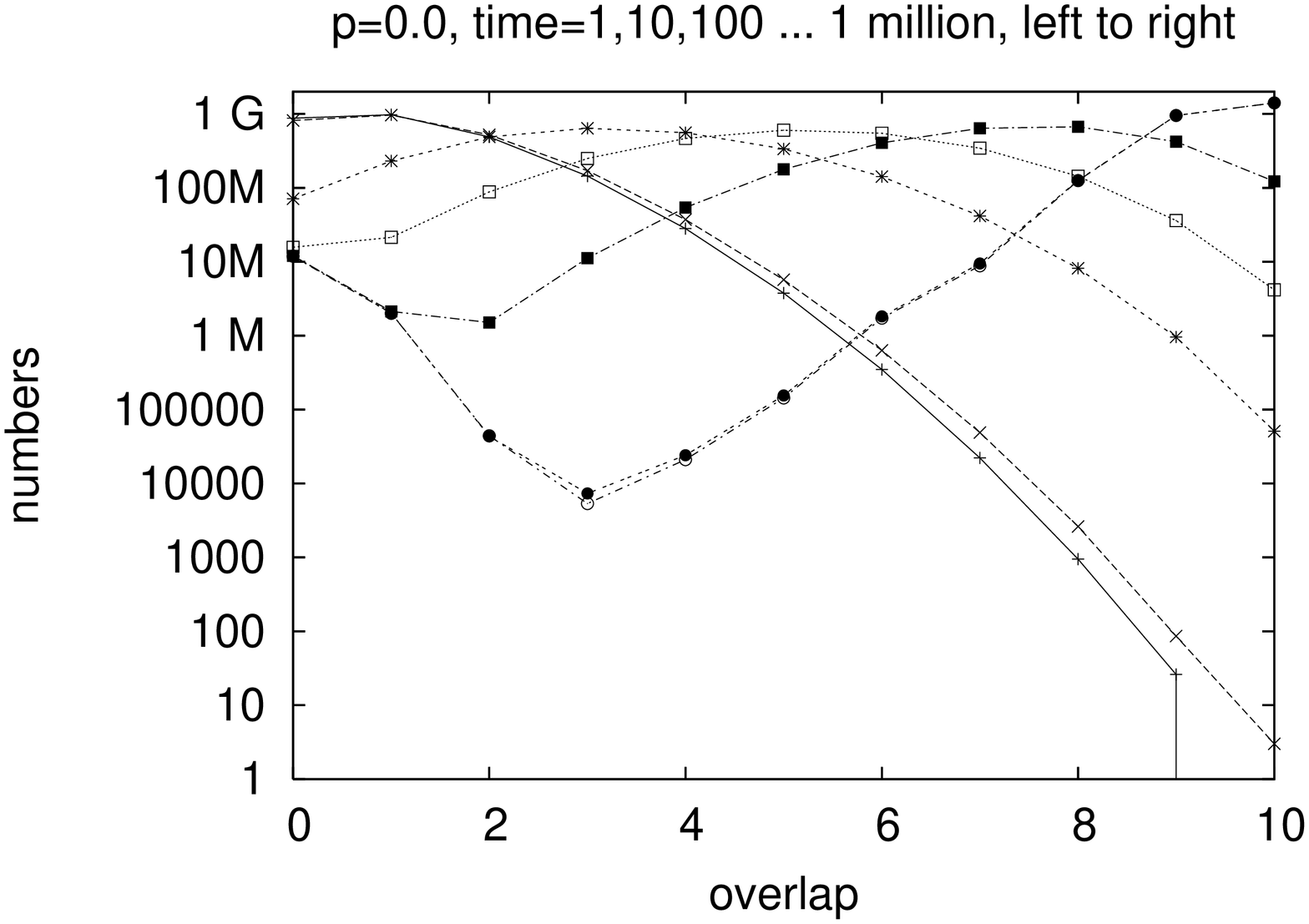}
\end{center}
\caption{Variation of overlap with observation time growing from 1 to $10^6$
in powers of ten; 5000 samples for 1000+3 sites, $\Delta = 30, \; p = 0.2$ and 0.0. 
}
\end{figure}  
%'axeldeff20a.fig' w lp, 'axeldeff20b.fig' w lp, 'axeldeff20c.fig' w lp, 'axeldeff20d.fig' w lp, 'axeldeff20e.fig' w lp, 'axeldeff20f.fig' w lp, 'axeldeff20.fig' w lp
% 'axeldeff24f.fig' w lp, 'axeldeff24e.fig' w lp, 'axeldeff24a.fig' w lp, 'axeldeff24b.fig' w lp, 'axeldeff24c.fig' w lp, 'axeldeff24d.fig' w lp, 'axeldeff17.fig' w lp

\begin{figure}[hbt]
\begin{center}
\includegraphics[angle=-90,scale=0.5]{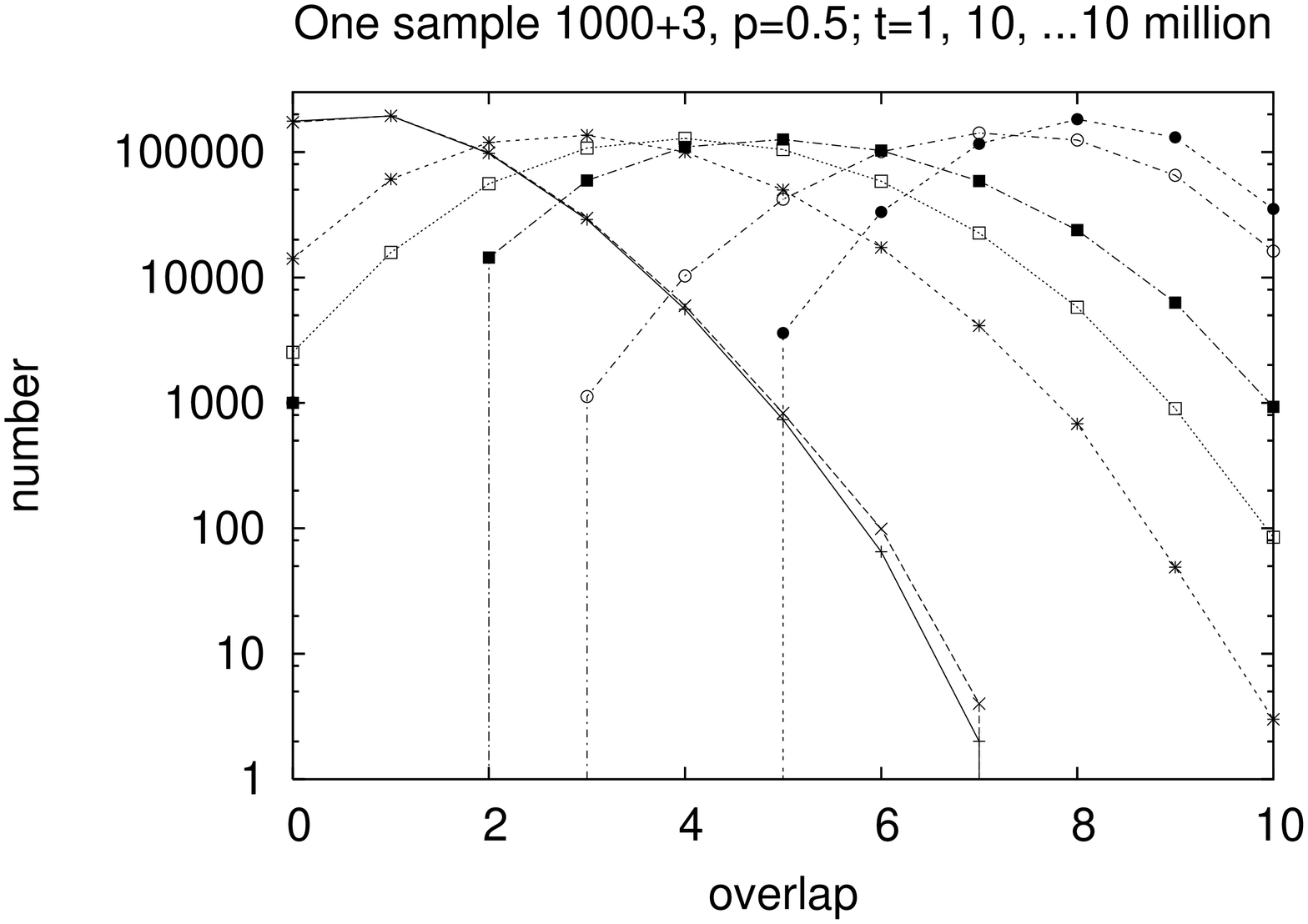}
\end{center}
\caption{Variation of overlap with observation time 1, 10,... 10 million; 
1000+3 sites, $\Delta = 30, \; p=0.5$.
}
\end{figure}  
% axeldeff26*.d, *=1 to g

Initially, the opinions are distributed randomly. Then, at each iterations, 
each person $i$ selects randomly a partner $j$. If
the total differences in the opinions $O$, summed over all ten themes, is not 
larger than some threshold $\Delta$, some interaction takes place.  Thus with 
opinions $O_{ik}$ of person $i$ on theme $k$, people $i$ and $j$ ignore each
other if $$\sum_k |O_{ik}-O_{jk}| > \Delta \quad .$$ An interaction means that
one of the ten themes is selected randomly; if both agree on this theme,
nothing happens. Otherwise, with probability $p$ person $i$ takes over the 
opinion of person $j$ (Axelrod case), while with probability $1-p$ they follow
the Deffuant opinion dynamics: 
$O_{ik}$ shifts towards $O_{jk}$ and $O_{jk}$ shifts towards
$O_{ik}$ by a (rounded) amount $\sqrt{0.1}|O_{jk}-O_{ik}|$. (If the two opinions
differ by only one unit, one of the two people takes over the opinion of the
other one.)

As in \cite{newbook} we summed over 5000 samples, each of which consisting
of 1000 agents surrounding a fully connected core of three people. Each agent
added to the Barab\'asi-Albert network selects three people to whom the 
new agent will seek contact later, and the selection probability for each of 
the three is proportional to the number of previous agents who had selected
these people as future contacts. The simulations were stopped when nobody 
changed opinion, but at the latest after one million iterations. 

Our Fig.1 shows that the previous results are not changed much if a small
probability $p$ to disobey Deffuant et al and to obey Axelrod is introduced:
Again most of the pairs of people either agree in the majority of the themes
or in none or one of them; seldomly there is agreement in three of the 
ten themes. Also the variation with $\Delta$ at $p=0.2$ in Fig.2 is about the 
same as at $p=0$, \cite{newbook}, and this is true also for 5000+3 sites
at $p = 0.1, \Delta = 30$.

In all these cases the simulations stopped before the maximum limit of $10^6$
updates per site for the observation time were reached. (For $p = 0.4$ and 0.9
this time was reached sometimes or always, respectively.) For opinion dynamics,
$10^6$ discussions for each person may be unrealistically many; Fig.3 shows
that for shorter times instead of a minimum we find a maximum at intermediate
overlaps. Again the results for $p=0$ and 0.2 are nearly the same.

All the above results sum over 1000 or 5000 samples; if only one sample is
followed one may see as in Fig.4 a Gaussian distribution shifting with
increasing time from agreement in very few to agreement in many themes.

In summary, the addition of a little bit Axelrod transfer to Deffuant opinion 
dynamics changed the results not much in these simulations.

We thank EU-FP6 grant MMCOMNET and the corresponding grant of the Polish 
Ministry of Science and Higher Education 13/6.PR UE/2005/7
for supporting the visit of DS.

\end{document}